\documentclass[UKenglish, autoref, runningheads]{llncs}

 \usepackage{color}
 \usepackage{xcolor}
 \usepackage{hyperref}
\usepackage{xurl}
\usepackage{todonotes}
\usepackage{listings}
\usepackage{booktabs}
\usepackage{tabularx}
\usepackage{multirow}

\usepackage{amsmath}
\usepackage{algorithm}
\usepackage{algpseudocode}

\usepackage{tikz}
\usepackage{fontawesome5}
\usetikzlibrary{arrows.meta,positioning,fit,backgrounds,calc}

\lstdefinestyle{figcode}{
  language=C,
  numbers=none,
  basicstyle=\ttfamily\fontsize{7.6pt}{8.6pt}\selectfont,
  mathescape=true,
  columns=fullflexible,
  keepspaces=true,
  breaklines=true,
  breakatwhitespace=true,
  xleftmargin=0pt,
  framexleftmargin=0pt,
  numbersep=0pt,
  aboveskip=0pt,
  belowskip=0pt,
  morekeywords=[2]{RANGE,COUNTER_OK, COUNTER_OVERFLOW},
  keywordstyle=[2]\color{purple}\bfseries,
}
\definecolor{contractgreen}{RGB}{2,112,10}
\definecolor{comment}{RGB}{0,128,0}

\usepackage{cleveref}

\newcommand{\inlineheadingbf}[1]{\medskip\noindent{\bfseries #1.}}
\newcommand{\inlineheadingit}[1]{\medskip\noindent{\it #1.}}

\title{Towards Agentic  Proof Harness Engineering \\ {\small (NIER paper)\vspace{-0.7em}}}
\author{Rohith Kumar Shanmuganathan$^1$ \and Cedric Richter$^2$ \and Heike Wehrheim$^1$}
 \institute{
	Carl von Ossietzky Universität Oldenburg, Oldenburg, Germany \and Humbold-Universität zu Berlin, Berlin, Germany\\
	\email{\{rohith.shanmuganathan, heike.wehrheim\}@uol.de \newline cedric.richter@hu-berlin.de}
}
\authorrunning{R.K.~Shanmuganathan, C.~Richter,  H.~Wehrheim}

\begin{document}

\maketitle
\vspace{-0.4em}
{\centering
\small
\textbf{Artifact:}
\href{https://doi.org/10.5281/zenodo.23378668}
     {https://doi.org/10.5281/zenodo.23378668}
\par}
\vspace{0.1em}

\begin{abstract}
Formal verification of real-world projects requires substantial manual effort, from specifying correctness properties to constructing the infrastructure needed to check them. We envision verification agents that assist developers throughout this process. Among the many tasks involved, one stands out as both foundational and amenable to automation: proof-harness engineering. A proof harness defines an isolated verification environment for a program unit, such as a function or module, by providing symbolic inputs, initialization code, and models of its dependencies. Today, proof-harness engineering remains largely a manual task performed by verification engineers. To explore the potential of coding agents for this task, we propose an agentic proof-harness engineer that automatically constructs verifier-ready harnesses from a software project and a designated verification target. We combine the agent with formal soundness and precision guards to validate generated abstractions and guide their refinement. We evaluate the feasibility of our approach in an exploratory case study on \textsc{curl} and \textsc{libuv}, using three historical assertion violations already embedded in the source code as verification properties. Our results show that the agent successfully constructs proof harnesses that enable CBMC to detect all three violations. We further demonstrate the potential for reusing these harnesses with other verification tools.
\end{abstract}

\section{Introduction} 
\label{sec:intro}

%
Automatic software verification has become a mature field, with numerous tools (\emph{verifiers}) supporting program verification. The annual Competition on Software Verification (SV-COMP)~\cite{DBLP:conf/tacas/BeyerS26} provides a large and diverse benchmark suite on which the effectiveness and efficiency of verifiers are evaluated. Despite significant progress in recent years, applying software verification tools to real-world software projects remains challenging. Verifiers typically operate on well-defined verification tasks, whereas real-world projects consist of multiple source files, depend on project-specific build configurations and external libraries, and require suitable initialization and environment models. Developers must therefore bridge the gap between a software project and a verifier-ready verification task.

In this paper, we envision \emph{coding agents} assisting developers in this process and focus on the task of \emph{proof-harness engineering}. A proof harness defines an isolated verification environment for a program unit, such as a function, by providing symbolic inputs, initialization code, and abstractions of its dependencies. Our goal is an autonomous verification-harness agent that constructs complete proof harnesses for C projects given the project source code, its build instructions, and a unit to be verified. The agent constructs the harness by recovering and abstracting information already present in the project, such as how the target is initialized, used, and interacts with its dependencies. The resulting harness should over-approximate the target's possible executions without excluding relevant program behaviors. Importantly, the agent does not invent the specification; the properties to be verified remain under the control of the developer. Such properties are frequently expressed in program code through assertions and sanity checks~\cite{DBLP:conf/icse/CasalnuovoDOFR15}.

Constructing a proof harness requires substantial project-specific knowledge. Recent work on \textsc{HarnessForge}~\cite{DBLP:conf/sigsoft/BeyerCHLL26} automates the extraction of self-contained verification tasks from C projects, but requires manually provided harness code, including symbolic inputs, initialization, and models of external dependencies. We investigate whether a coding agent can autonomously construct the complete proof harness from the project itself, without requiring developers to provide any harness code. To address the risk of unsound or imprecise abstractions, we introduce independent \emph{formal guardrails} that check the soundness of dependency abstractions and guide their refinement using verifier feedback.

To evaluate the feasibility of agentic proof-harness engineering, we implement a prototype and conduct an exploratory case study on two real-world C projects, \textsc{curl} and \textsc{libuv}. We use three historical assertion violations as verification targets. Our agent successfully constructs proof harnesses that enable CBMC to detect all three violations. We further investigate the reuse of these harnesses with other verification tools, demonstrating their potential beyond the verifier used during construction while also exposing limitations in cross-verifier compatibility.







\section{Agentic Proof Harness Engineering}
\label{sec:harness} 
\begin{figure}[t]
\begin{minipage}[t]{.6\textwidth}
\begin{lstlisting}[style=figcode]
// Target function in "./src/counter.c"
counter_status_t add(short some_value) {
    short value = toAdd(some_value);

    if (...){
        // Over-flow handling
        return COUNTER_OVERFLOW; 
    }

    short noi_old = gc.noi;
    short cv_old  = gc.cv;

    gc.noi++;
    gc.cv += value;

    short diff = gc.cv - cv_old;
    assert(gc.noi >= noi_old &&
            0 <= diff && diff <= RANGE);

    return COUNTER_OK;
}
\end{lstlisting}
\end{minipage}
\hfill
\begin{minipage}[t]{.6\textwidth}
\begin{lstlisting}[style=figcode]
// Proof Harness
#include "./src/counter.c"

short toAdd(short val) {
    short r = nondet_short();
    assume(r >= 0 && r <= RANGE 
        $||$ (val == SHRT_MIN 
              && r == -RANGE));
    return r;
}

int main(void) {
    gc.noi = nondet_short();
    gc.cv = nondet_short();
    add(nondet_short());
    return 0;
}
\end{lstlisting}
\end{minipage}
\caption{Target function (\emph{left}) and a corresponding proof harness (\emph{right}).}
\label{fig:example}
\end{figure} 
Our goal is to provide initial insights on how (coding) agents can be employed to automate the task of proof harness engineering. We demonstrate the task with a simple example taken and adapted from Beyer et al.~\cite{DBLP:conf/sigsoft/BeyerCHLL26}\footnote{\url{https://gitlab.com/sosy-lab/software/harnessforge/-/tree/main/doc/tutorial}}. 

\inlineheadingbf{Proof Harness Engineering} \Cref{fig:example} (left) shows an example function \texttt{add} implemented as part of a larger software system. Our goal is to automatically check whether the assertion encoded in the function can be violated. Verifying this property directly in the context of the complete system is challenging: \texttt{add} may be reachable from thousands of call sites, while the states in which it is invoked depend on the composition, configuration, and environment of the project.

A proof harness provides a compact abstraction of the execution environment. \Cref{fig:example} (right) shows a corresponding harness for \texttt{add}. Instead of enumerating its callers, the harness initializes the relevant global state symbolically and invokes \texttt{add} with a symbolic input. The dependency \texttt{toAdd} is replaced by a model that summarizes its relevant behavior. The verifier can then analyze the resulting self-contained task independently of the remainder of the project.

Proof-harness engineering requires substantial project-specific knowledge. The engineer must determine which program state influences the target and how it should be initialized, replace omitted dependencies by abstractions that preserve their relevant behavior, and account for the project's build configuration, including types, macros, and other compile-time choices. In our example, the abstraction of \texttt{toAdd} encodes a formal \emph{summary} relating its inputs to possible return values. In particular, for \texttt{some\_value == SHRT\_MIN}, the dependency may return \texttt{-RANGE}, which can lead to an assertion violation in \texttt{add}. Omitting this behavior from the abstraction could therefore hide a real violation. Constructing such a harness therefore requires exploring and reasoning about information distributed throughout the project.

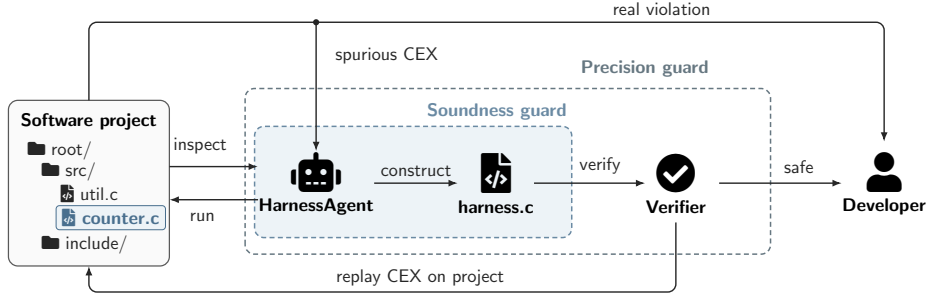
\begin{figure}[t]
\centering
\resizebox{\linewidth}{!}{%
\begin{tikzpicture}[
  x=1mm, y=1mm, >=Latex,
  font=\sffamily,
  flow/.style={-{Latex[length=2mm]}, line width=.85pt, draw=black!85},
  plain/.style={line width=.85pt, draw=black!85},
  note/.style={font=\sffamily\fontsize{10.5}{12}\selectfont, inner sep=0pt, text=black!85},
  tag/.style={font=\sffamily\bfseries\fontsize{11}{12}\selectfont, inner sep=0pt},
  item/.style={font=\sffamily\bfseries\fontsize{11}{12}\selectfont, inner sep=0pt, align=center},
  region/.style={rounded corners=2mm, line width=.9pt, densely dashed}
]
\definecolor{guardLine}{HTML}{6A8AA3}
\definecolor{guardFill}{HTML}{EDF3F7}
\definecolor{refineLine}{HTML}{667781}
\definecolor{targetLine}{HTML}{496F8E}

\coordinate (P) at (17,0);
\coordinate (A) at (65,0);
\coordinate (H) at (103,0);
\coordinate (V) at (141,0);
\coordinate (U) at (185,0);

\begin{scope}[on background layer]
  \draw[region,draw=refineLine] (50,-15) rectangle (161,20);
  \filldraw[region,draw=guardLine,fill=guardFill] (52,-11.5) rectangle (119,12);
\end{scope}
\node[tag,text=guardLine,anchor=east] at (118,15.2) {Soundness guard};
\node[tag,text=refineLine,anchor=east] at (148,24.2) {Precision guard};

\draw[rounded corners=2mm,draw=black!65,fill=black!2,line width=.75pt]
  (0,-17.5) rectangle (34,17.5);
\node[tag] at (17,13) {Software project};
\node[note,anchor=west] at (4,7) {\faFolder\ root/};
\node[note,anchor=west] at (7,2.4) {\faFolder\ src/};
\node[note,anchor=west] at (11,-2.2) {\faFileCode\ util.c};
\node[note,anchor=west,draw=targetLine,fill=guardFill,
      rounded corners=.7mm,inner xsep=1.3mm,inner ysep=.7mm,
      text=targetLine,font=\sffamily\bfseries\fontsize{10.5}{12}\selectfont]
      at (10,-7.3) {\faFileCode\ counter.c};
\node[note,anchor=west] at (7,-13) {\faFolder\ include/};

\node[item] (agent) at (A) {{\fontsize{24}{25}\selectfont\faRobot}\\[1mm]HarnessAgent};
\node[item] (harness) at (H) {{\fontsize{24}{25}\selectfont\faFileCode}\\[1mm]harness.c};
\node[item] (verifier) at (V) {{\fontsize{24}{25}\selectfont\faCheckCircle}\\[1mm]Verifier};
\node[item] (developer) at (U) {{\fontsize{24}{25}\selectfont\faUser}\\[1mm]Developer};

\draw[flow] (agent.east) -- node[note,above=1.3mm] {construct} (harness.west);
\draw[flow] (harness.east) -- (verifier.west);
\node[note,anchor=south] at (125,2) {verify};
\draw[flow] ([xshift=2.7mm]verifier.east) -- (developer.west);
\node[note,anchor=south] at (167,2) {safe};

\draw[flow] (34,3.5) -- (agent.west |- -1,3.5);
\node[note,anchor=south] at (40.5,6.2) {inspect};
\draw[flow] (agent.west |- -1,-3.5) -- (34,-3.5);
\node[note,anchor=north] at (41.0,-6.2) {run};

\draw[flow,rounded corners=1.7mm]
  ($(verifier.south) + (0, -2)$) -- (141,-23) -- (17,-23) -- (17,-17.5);
\node[note,anchor=south] at (87,-21.7) {replay CEX on project};

\draw[plain,rounded corners=1.7mm] (17,17.5) -- (17,34) -- (65,34);
\fill[black!85] (65,34) circle[radius=.65mm];

\draw[flow] (65,34) -- (agent.north);
\node[note,anchor=west] at (69,27.0) {spurious CEX};

\draw[flow,rounded corners=1.7mm] (65,34) -- (185,34) -- ($(developer.north) + (0, 2)$);
\node[note,anchor=south] at (138,35.5) {real violation};
\end{tikzpicture}%
}
\caption{Workflow of Agentic Proof Harness Engineering}
\label{fig:overview}
\end{figure}

\inlineheadingbf{Agentic Engineering}
Given a target system $S$, a function $f$, and a verifier $\mathcal{V}$, we seek to construct a harness $H_f$ that satisfies two requirements:
\begin{itemize}
\item \textbf{Soundness.} The harness over-approximates the inputs, program states, and dependency behaviours that may occur when executing $f$ in $S$. 
\item \textbf{$\mathcal{V}$-precision.} The over-approximation is sufficiently precise for $\mathcal{V}$ to establish the property encoded in $f$ or expose a feasible counterexample.
\end{itemize}
Constructing such a harness requires exploring the system $S$, identifying dependencies of $f$, recovering relevant project information, and constructing formal abstractions. We envision an autonomous agent performing this engineering work by inspecting the project and its configurations, executing tools, and iteratively constructing and revising candidate harnesses. \Cref{fig:overview} illustrates this workflow. 

Since the agent itself provides no formal guarantees, we propose two formal guardrails. A \emph{soundness guard} (\Cref{alg:harness-agent}) checks whether each abstraction $A_d$ of a dependency $d$ over-approximates the behavior of the original dependency. Violations are returned to the agent for repair. A precision guard (\Cref{alg:proof-harness}) uses feedback from $\mathcal{V}$ to identify insufficiently precise abstractions and guide their refinement.  Failed checks and spurious counterexamples are returned to the agent for revision. A more detailed description is provided in \Cref{app:setup}.


\inlineheadingbf{Related Work}
Agentic approaches to software engineering are rapidly emerging, including agents for general software-engineering tasks~\cite{yang2024sweagent}, program repair~\cite{11029914}, and automated program analysis~\cite{DBLP:journals/corr/abs-2604-11270}. Their application to formal verification is even more recent, with work on agentic interaction with proof assistants~\cite{Tu_2026,sosso2026agenticprovingprogramverification} and specification and proof generation~\cite{ding2026fmagentscalingformalmethods}. Closest to our proposal, agentic model checking~\cite{sun2026agenticmodelchecking} employs agents to synthesize specifications, stubs, and function-level verification code for bounded model checking. Our approach focuses on the construction of verification environments from existing project information. The verification properties remain developer-specified, and the generated harnesses can be processed by existing verifiers through an SV-COMP-compatible representation. Prior work on proof-harness construction~\cite{DBLP:journals/spe/ChongCEKKMSTTT21,DBLP:conf/tacas/BeyerCHLL26,DBLP:conf/sigsoft/BeyerCHLL26} automates parts of this process but still requires manually supplied harness code.

\section{Evaluation}
\begin{table}[t]
\centering
\caption{Historical assertion violations selected for
our case study.}
\label{tab:bugs}
\small
\begin{tabularx}{\linewidth}{@{}lclX@{}}
\toprule
\textbf{Project} & \textbf{PR} & \textbf{Component} & \textbf{Bug Description} \\
\midrule
\textsc{curl} & \href{https://github.com/curl/curl/pull/22785}{\#22785}
& DNS-over-HTTPS
& An assertion in \texttt{doh\_resp\_decode\_httpsrr()}
  fails on malformed DNS responses containing trailing bytes. \\
\addlinespace
\textsc{curl} & \href{https://github.com/curl/curl/pull/23292}{\#23292}
& HTTP headers
& Malformed headers with empty names violate an assertion
  in \texttt{Curl\_dynhds\_add()}. \\
\addlinespace
\textsc{libuv} & \href{https://github.com/libuv/libuv/pull/5021}{\#5021}
& Unicode
& An off-by-one assertion in \texttt{uv\_wtf8\_to\_utf16()}
  rejects the valid Unicode code point \texttt{0x10FFFF}. \\
\addlinespace
\Cref{fig:example} & --
& Counting
& \texttt{toAdd()} overflows for \texttt{SHRT\_MIN} \\
\bottomrule
\end{tabularx}
\end{table}
To evaluate the feasibility of agentic proof-harness engineering, we implement a prototype of the proposed harness agent. We conduct an exploratory case study on \textsc{curl}\footnote{\url{https://curl.se/}}, a widely used C project comprising more than 4,500 files and 630,000 lines of code\footnote{\url{https://curl.se/gitstats/}}. We investigate whether our agent can construct proof harnesses to expose two historical bugs in \textsc{curl} and additionally explore its applicability to a historical bug in \textsc{libuv}\footnote{\url{https://github.com/libuv/libuv}} and our example in \Cref{fig:example}. Our evaluation is guided by the following research question:

\medskip
\noindent Can an agent construct proof harnesses that expose and verify historical bugs?
\medskip

\subsection{Experimental Setup}
We select three historical assertion violations documented in pull requests from \textsc{curl} and \textsc{libuv}: \textsc{curl}~\#22785, \textsc{curl}~\#23292, and \textsc{libuv}~\#5021. \Cref{tab:bugs} provides an overview of the selected cases. We selected these pull requests because (1) they cover different functionalities within the respective projects, (2) the violated assertions were already present in the source code, allowing us to focus on the agent's harness construction capabilities, and (3) the corresponding pull requests provide the original and fixed revisions for evaluating the generated harnesses. We also discuss our example with an assertion violation from \Cref{fig:example}. The agent is employed to build a harness for the original and fixed revisions independently. We provide the agent with the name of target function in its initial prompt. The prompt is shown in \Cref{sec:prompt}. We use \texttt{openai/gpt-6-luna} with a temperature of 0.0 and limit of 60 actions per run.


To avoid interference with the original project, we isolate agent runs: The agent can read every file in the original project, but can only modify and create files in a dedicated \texttt{verification/} directory. Actions are executed in an isolated Docker container. At the end of the agent run, the \texttt{verification/} directory should include a proof harness that compiles against the original project. Our main verifier is CBMC~\texttt{v5.95.1}\footnote{\url{https://www.cprover.org/cbmc/}} with a timeout of 5 minutes\footnote{Since the agent might call CBMC multiple times inside the loop, each verifier call is limited to 5 minutes to keep feedback fast.}.  The harness follows the SV-COMP conventions, thus allowing us to employ different verifiers to the same harness. We also evaluated UAutomizer~\cite{DBLP:conf/tacas/SchusseleBDHJKP24} (the winner of the SV-COMP 2026), CPAchecker~\cite{CPAchecker}, and KLEE~\cite{DBLP:journals/sttt/CadarN21} on the harnesses generated for CBMC. 
All experiments were carried out on a single machine equipped with an Intel Xeon W-2265 CPU (12 physical cores, 24 hardware threads, 4.8 GHz) and 251 GiB of RAM, running Ubuntu 22.04.

\subsection{Results}

\begin{table}[t]
  \centering
  \caption{Verification results for historical assertion
    violations. \faCheck: safe; \faTimes: unsafe;
    ?: unknown; E: error.
    Desired outcomes are \faTimes~for violated assertions
    and \faCheck~or ? for resolved assertions
    (the latter remains inconclusive).}
  \label{tab:results}

  \small
  \setlength{\tabcolsep}{4pt}

  \begin{tabular}{@{}llc@{\hspace{9pt}}cccc@{}}
    \toprule
    & & \multicolumn{1}{c}{\textbf{HarnessAgent}}
        && \multicolumn{3}{c}{\textbf{Harness Reuse}} \\
    \cmidrule(lr){3-3}
    \cmidrule(l){4-7}
    Case & Assertion & \textbf{CBMC}
        && KLEE & UAutomizer & CPAchecker \\
    \midrule
    \textsc{curl} \#22785
      & Violated & \faTimes && \faTimes & E & E \\
      & Resolved & ? && ? & E & E \\
    \addlinespace
    \textsc{curl} \#23292
      & Violated & \faTimes && \faTimes & E & ? \\
      & Resolved & ? && ? & \faCheck & ? \\
    \addlinespace
    \textsc{libuv} \#5021
      & Violated & \faTimes && \faTimes & \faTimes & \faTimes \\
      & Resolved & \faCheck && ? & ? & \faCheck \\
      \addlinespace
    \Cref{fig:example}
      & Violated & \faTimes && \faTimes & \faTimes & \faTimes \\
    \bottomrule
  \end{tabular}
\end{table}

In \Cref{tab:results}, we report the results of our harness construction using CBMC as the guiding verifier. We additionally evaluate the generated harnesses with three other verifiers. For all three pull requests (cases), our harness agent successfully constructs a proof harness that enables CBMC to detect the historical assertion violation in the original revision. When applied to the revised versions, CBMC no longer reports a violation, returning \faCheck{} (safe) or ? (unknown). Notably, CBMC establishes safety for the revised version of \textsc{libuv}~\#5021. In this case, the fix corrects an assertion on the maximum Unicode code point, making it consistent with the range accepted by the decoder.

\inlineheadingit{Harness reuse}
Proof harnesses are not necessarily one-time artifacts; they offer two opportunities for reuse. First, developers can reuse proof harnesses across revisions of a project, for example, to reverify an implementation after fixing an assertion violation. With CBMC, we found that two of the three generated harnesses could be reused without modification, while one required further refinement of the dependency abstraction. Second, developers can reuse proof harnesses with different verifiers. We investigate this scenario by analyzing the generated harnesses with KLEE, Ultimate Automizer, and CPAchecker.
For \textsc{libuv}, all four verifiers detect the historical assertion violation. On the resolved revision, CBMC and CPAchecker establish safety, while KLEE and Ultimate Automizer return unknown. For \textsc{curl}, CPAchecker and Ultimate Automizer struggle with unsupported C constructs and library functions, potentially due to differences between the project's C dialect and the GNU C subset supported by SV-COMP verifiers. KLEE requires additional definitions from \textsc{curl} source files, which we provide manually. With these additions, KLEE detects both historical assertion violations in \textsc{curl} and reports unknown on the resolved revisions.

These results provide initial evidence that agent-generated proof harnesses can be reused across project revisions and verification tools, while highlighting the challenges of achieving portability without manual adaptations.
%
%

\inlineheadingit{Impact of Guards} 
We investigate the impact of our soundness and precision guards on harness construction for the motivating example in \Cref{fig:example} (\Cref{tab:results}, last row). We find that while \texttt{openai/gpt-6-luna} is capable of constructing sound harnesses, it struggles to achieve sufficient precision to distinguish real assertion violations from spurious counterexamples. Our precision guard identifies spurious counterexamples and guides the agent toward increasingly precise harnesses, while the soundness guard ensures that refinements preserve the behavior of the original dependencies. A qualitative analysis of the refinement process is provided in \Cref{sec:refinement}. The precision guard also proves valuable in our real-world case studies, preventing 11 false-positive reports with CBMC.

\section{Discussion and Future Research}
We discuss two research directions arising from our approach and the remaining challenges toward \emph{trustworthy agentic system verification}.

\inlineheadingbf{Towards Trustworthy Agentic System Verification}
We envision coding agents assisting developers throughout the verification process, from constructing proof harnesses for components of large software systems to configuring verification tools, interpreting results, and repairing violations. Today's coding agents are already capable of assisting with many of these tasks, but they provide no correctness guarantees. An agent may fail unpredictably, incorrectly claim that a program has been verified, or modify the specification to achieve a desired verification result. Such behavior undermines trust in agent-generated results, particularly in safety-critical software development.

We believe that our approach provides a first step toward \emph{trustworthy agentic system verification}. By combining coding agents with independent formal guardrails, we aim to establish trust in agent-generated artifacts without relying on the correctness of the agents' reasoning. A natural next step is to extend proof-harness engineering from individual functions to entire software systems. We envision agents systematically constructing verification environments for program units throughout a project, exploiting information about their interactions to improve precision and enable system-level reasoning. Establishing formal guarantees when combining such local analyses remains a key research challenge.

\inlineheadingbf{Towards Automated Benchmark Construction}
Constructing software verification benchmarks is challenging and time-consuming. Existing benchmark suites contain artificially generated programs and tasks extracted from real-world projects, for which dependencies and execution environments often require manual abstraction. Agentic proof-harness engineering could automate this process, enabling larger and more diverse benchmark suites from real-world software. A central challenge is ensuring that generated tasks preserve the properties of the original software. While sound over-approximations support safety proofs, they may introduce spurious violations. This raises the question of how to construct sufficiently precise tasks with expected verdicts that are independent of the verifier used during construction. In the longer term, if proof-harness engineering becomes an integral part of system-level verification, the resulting benchmarks could directly reflect the workloads encountered by verifiers in practice.

\inlineheadingbf{Challenges of Proof-Harness Engineering}
Several challenges for agentic proof-harness engineering remain. First, validating dependency abstractions may be infeasible when implementations are unavailable or verification is inconclusive. Dynamic analysis could help construct candidate abstractions in such cases, although establishing soundness remains a challenge. Second, isolated verification environments may admit initial states unreachable in the original system. Recovering relevant execution contexts while preserving developer-defined specifications remains an open challenge. Finally, our current refinement strategy provides no guarantee of progress or termination, as the agent may repeatedly fail to construct sufficiently precise abstractions.

\inlineheadingbf{Limitations of Coding Agents}
Our experiments also exposed practical limitations of coding agents. Agents entered unproductive loops, spent considerable effort searching for suitable build configurations, and attempted prohibited operations, including modifying developer-provided assertions. Restricting project access mitigates some of these problems. Agent runs are also nondeterministic. In two runs on a \textsc{curl} case, we observed different strategies: in the first, the agent identified the violation through code reasoning and confirmed it through concrete execution; in the second, CBMC discovered the violation directly. Such variability complicates reproducibility and motivates further research into controlling and evaluating agentic verification workflows.
\section{Conclusion}
In this paper, we propose the problem of \emph{agentic proof-harness engineering} and present our first steps toward addressing it. Our prototype combines a coding agent with independent soundness and precision guards to construct verification tasks from real-world C projects. An exploratory case study on \textsc{curl} and \textsc{libuv} demonstrates the feasibility of this approach, enabling existing verifiers to expose historical assertion violations. Our results also provide initial evidence that generated proof harnesses can be reused across different verifiers. We see agentic proof-harness engineering as a promising first step toward \emph{trustworthy agentic system verification}, with opportunities for system-level verification and automated benchmark construction.



\bibliographystyle{plainurl}

\bibliography{lit}

@inproceedings{DBLP:conf/tacas/BeyerS26,
  author       = {Dirk Beyer and
                  Jan Strejcek},
  editor       = {Sebastian Junges and
                  Guy Katz},
  title        = {{Evaluating Software Verifiers for C, Java, and {SV-LIB} - (Report
                  on {SV-COMP} 2026)}},
  booktitle    = { {TACAS} },
  series       = {Lecture Notes in Computer Science},
  volume       = {16506},
  pages        = {461--502},
  publisher    = {Springer},
  year         = {2026},
  doi          = {10.1007/978-3-032-22749-2\_23},
  bibsource    = {dblp computer science bibliography, https://dblp.org}
}

@article{DBLP:journals/sttt/CadarN21,
  author       = {Cristian Cadar and
                  Martin Nowack},
  title        = {{KLEE} symbolic execution engine in 2019},
  journal      = {Int. J. Softw. Tools Technol. Transf.},
  volume       = {23},
  number       = {6},
  pages        = {867--870},
  year         = {2021},
  url          = {https://doi.org/10.1007/s10009-020-00570-3},
  doi          = {10.1007/S10009-020-00570-3},
  bibsource    = {dblp computer science bibliography, https://dblp.org}
}

@inproceedings{DBLP:conf/tacas/SchusseleBDHJKP24,
  author       = {Frank Sch{\"{u}}ssele and
                  Manuel Bentele and
                  Daniel Dietsch and
                  Matthias Heizmann and
                  Xinyu Jiang and
                  Dominik Klumpp and
                  Andreas Podelski},
  title        = {{Ultimate Automizer and the Abstraction of Bitwise Operations - (Competition
                  Contribution)}},
  booktitle    = {{TACAS} {(3)}},
  series       = {Lecture Notes in Computer Science},
  volume       = {14572},
  pages        = {418--423},
doi          = {10.1007/978-3-031-57256-2\_31},
  publisher    = {Springer},
  year         = {2024}
}

@InProceedings{CPAchecker,
  author =     {Dirk Beyer and M.~Erkan~Keremoglu},
  title =      {{\textsc{CPAchecker}: A Tool for Configurable Software Verification}},
  booktitle =  {Proc.\ CAV},
  publisher =  {Springer},
  series =     {LNCS~6806},
  pages =      {184--190},
  year =       {2011},
  doi  =       {10.1007/978-3-642-22110-1\_16},
}

@inproceedings{yang2024sweagent,
  title={{SWE}-agent: Agent-Computer Interfaces Enable Automated Software Engineering},
  author={John Yang and Carlos E Jimenez and Alexander Wettig and Kilian Lieret and Shunyu Yao and Karthik R Narasimhan and Ofir Press},
  booktitle={The Thirty-eighth Annual Conference on Neural Information Processing Systems},
  year={2024},
  url={https://arxiv.org/abs/2405.15793}
}

@article{sun2026agenticmodelchecking,
      title={{Agentic Model Checking}}, 
      author={Youcheng Sun and Jiawen Liu and Daniel Kroening and Jason Xue},
      year={2026},
      journal      = {CoRR},
  volume       = {abs/2605.21434},
      eprint={2605.21434},
        eprinttype   = {arXiv},
      primaryClass={cs.SE},
      url={https://arxiv.org/abs/2605.21434}, 
}

@article{ding2026fmagentscalingformalmethods,
      title={{FM-Agent: Scaling Formal Methods to Large Systems via LLM-Based Hoare-Style Reasoning}}, 
      author={Haoran Ding and Zhaoguo Wang and Haibo Chen},
      year={2026},
      journal      = {CoRR},
  volume       = {abs/2604.11556},
      eprint={2604.11556},
      archivePrefix={arXiv},
      primaryClass={cs.SE},
      url={https://arxiv.org/abs/2604.11556}, 
}

@article{sosso2026agenticprovingprogramverification,
      title={Agentic Proving for Program Verification}, 
      author={Alessandro Sosso and Akhil Arora and Bas Spitters},
      year={2026},
      eprint={2605.23772},
      journal      = {CoRR},
  volume       = {abs/2605.23772},
      archivePrefix={arXiv},
      primaryClass={cs.AI},
      url={https://arxiv.org/abs/2605.23772}, 
}

@article{Tu_2026,
   title={Agentic Verification of Software Systems},
   volume={3},
   ISSN={2994-970X},
   url={http://dx.doi.org/10.1145/3808164},
   DOI={10.1145/3808164},
   number={FSE},
   journal={Proceedings of the ACM on Software Engineering},
   publisher={Association for Computing Machinery (ACM)},
   author={Tu, Haoxin and Zhao, Huan and Song, Yahui and Zafar, Mehtab and Meng, Ruijie and Roychoudhury, Abhik},
   year={2026},
   month=June, pages={3558–3581} }

@INPROCEEDINGS{11029914,
  author={Bouzenia, Islem and Devanbu, Premkumar and Pradel, Michael},
  booktitle={  ICSE}, 
  title={{RepairAgent: An Autonomous, LLM-Based Agent for Program Repair}}, 
  year={2025},
  volume={},
  number={},
  pages={2188-2200},
  doi={10.1109/ICSE55347.2025.00157}}

@inproceedings{DBLP:conf/tacas/BeyerCHLL26,
  author       = {Dirk Beyer and
                  Po{-}Chun Chien and
                  Bo{-}Yuan Huang and
                  Nian{-}Ze Lee and
                  Thomas Lemberger},
  editor       = {Sebastian Junges and
                  Guy Katz},
  title        = {A Case Study in Firmware Verification: Applying Formal Methods to
                  {Intel TDX}
                  Module},
  booktitle    = {  {TACAS}  , Part {II}},
  series       = {Lecture Notes in Computer Science},
  volume       = {16506},
  pages        = {42--64},
  publisher    = {Springer},
  year         = {2026},
  url          = {https://doi.org/10.1007/978-3-032-22749-2\_3},
  doi          = {10.1007/978-3-032-22749-2\_3},
  bibsource    = {dblp computer science bibliography, https://dblp.org}
}

@article{DBLP:journals/corr/abs-2604-11270,
  author       = {Islem Bouzenia and
                  Cristian Cadar and
                  Michael Pradel},
  title        = {Evaluating {LLM} Agents on Automated Software Analysis Tasks},
  journal      = {CoRR},
  volume       = {abs/2604.11270},
  year         = {2026},
  url          = {https://doi.org/10.48550/arXiv.2604.11270},
  doi          = {10.48550/ARXIV.2604.11270},
  eprinttype   = {arXiv},
  eprint       = {2604.11270},
  bibsource    = {dblp computer science bibliography, https://dblp.org}
}

@inproceedings{DBLP:conf/sigsoft/BeyerCHLL26,
  author       = {Dirk Beyer and
                  Po{-}Chun Chien and
                  Bo{-}Yuan Huang and
                  Nian{-}Ze Lee and
                  Thomas Lemberger},
  title        = {{HarnessForge: Automated Extraction of Verification Tasks from Industry-Scale
                  Software Projects}},
  booktitle    = {  {FSE} Companion },
  pages        = {202--206},
  publisher    = {{ACM}},
  year         = {2026},
  url          = {https://doi.org/10.1145/3803437.3806420},
  doi          = {10.1145/3803437.3806420},
  bibsource    = {dblp computer science bibliography, https://dblp.org}
}

@article{DBLP:journals/spe/ChongCEKKMSTTT21,
  author       = {Nathan Chong and
                  Byron Cook and
                  Jonathan Eidelman and
                  Konstantinos Kallas and
                  Kareem Khazem and
                  Felipe R. Monteiro and
                  Daniel Schwartz{-}Narbonne and
                  Serdar Tasiran and
                  Michael Tautschnig and
                  Mark R. Tuttle},
  title        = {{Code-level model checking in the software development workflow at
                  Amazon Web Services}},
  journal      = {Softw. Pract. Exp.},
  volume       = {51},
  number       = {4},
  pages        = {772--797},
  year         = {2021},
  url          = {https://doi.org/10.1002/spe.2949},
  doi          = {10.1002/SPE.2949},
  bibsource    = {dblp computer science bibliography, https://dblp.org}
}

@inproceedings{DBLP:conf/icse/CasalnuovoDOFR15,
  author       = {Casey Casalnuovo and
                  Premkumar T. Devanbu and
                  Ab{\'{\i}}lio Oliveira and
                  Vladimir Filkov and
                  Baishakhi Ray},
  editor       = {Antonia Bertolino and
                  Gerardo Canfora and
                  Sebastian G. Elbaum},
  title        = {Assert Use in GitHub Projects},
  booktitle    = {37th {IEEE/ACM} International Conference on Software Engineering,
                  {ICSE} 2015, Florence, Italy, May 16-24, 2015, Volume 1},
  pages        = {755--766},
  publisher    = {{IEEE} Computer Society},
  year         = {2015},
  url          = {https://doi.org/10.1109/ICSE.2015.88},
  doi          = {10.1109/ICSE.2015.88},
  bibsource    = {dblp computer science bibliography, https://dblp.org}
}

\clearpage
\appendix
\section{Agentic Proof Harness Engineering: Details}
\label{app:setup}

We implement a prototype of the agentic proof-harness engineer for C projects introduced in \Cref{sec:harness}. Our design consists of a harness agent guided by two components: a soundness guard that validates dependency abstractions and a precision guard that uses verifier feedback to guide their refinement. The soundness guard is described in \Cref{alg:harness-agent} and the precision guard in \Cref{alg:proof-harness}. Importantly, the agent does not need to construct a sufficiently precise harness in a single step. It starts with a permissive abstraction and progressively improves its precision, while each restriction on dependency behavior is checked for soundness.

\subsection{Harness Agent}
We implemented a custom LLM-based agent to construct a proof harness $H_f$ from a given C project $S$ and target function $f$. Our design goal was to construct \emph{over-approximating} proof harnesses without modifying the original project $S$.

\inlineheadingit{Proof Harness Initialization}
The agent starts by locating the target function in the C project, inspecting its signature, identifying its dependencies, and searching for a suitable build configuration. It then constructs a proof harness similar to \Cref{fig:example}. First, the agent creates a \emph{driver function} \texttt{main} that initializes the inputs and global state used by the target function and invokes it. To over-approximate the possible initial states, the driver assigns nondeterministic values to the inputs and relevant global variables. We deliberately impose no additional restrictions on these values, treating the target function as an independently invoked unit. Second, the agent constructs symbolic mocks ({\em abstractions}) for dependencies that are excluded from the verification task. Initially, these abstractions  admit unconstrained behavior. Their precision is subsequently improved through verifier-guided refinement, with each proposed restriction checked by the soundness guard.

\inlineheadingit{Restricted Project Access}
During proof-harness engineering, the agent has read-only access to the original C project $S$ and may modify files only within a dedicated \texttt{verification/} directory. We enforce these restrictions at the operation level by blocking unauthorized file operations and commands and redirecting the agent to permitted operations. We additionally prohibit access to the project's Git history. We found these restrictions necessary, as the agent attempted to inspect previous commits and modify developer-provided assertions to achieve its verification goal. By restricting these operations, we ensure that the agent constructs the harness without altering the verification problem.

\inlineheadingit{Tools}
To support proof-harness engineering, we provide the agent with the following tools:
\begin{itemize}
\item \texttt{describe\_schema}: Describes the expected proof-harness format.
\item \texttt{verify}: Executes CBMC on the current harness and checks reported counterexamples against the original implementation. It reports \emph{safe}, \emph{spurious counterexample}, or \emph{reproduced violation}.
\item \texttt{validate}: Checks whether the current dependency abstractions over-approximate the behavior of their original implementations. It reports \emph{sound} or provides a counterexample demonstrating an unsound abstraction.
\item \texttt{test}: Executes the original implementation with concrete inputs chosen by the agent.
\end{itemize}
These tools allow the agent to perform the soundness and precision checks described in our approach. However, the availability of these tools does not guarantee that the agent invokes them correctly or sufficiently, nor that the resulting harness is sound and $\mathcal{V}$-precise. We therefore employ external soundness and precision guards that enforce validation independently of the agent's decisions and invoke further refinement when necessary.

\inlineheadingit{Configuration}
\begin{table}[t]
  \centering
  \caption{Agent configuration.}
  \label{tab:config}
  \begin{tabular}{l @{\hspace{1.5em}} | @{\hspace{1.5em}} l}
    \toprule
    Parameter & Value \\
    \midrule
    Model & \texttt{gpt-6-luna} (via LiteLLM) \\
    Reasoning effort & off \\
    Temperature & 0; 0.7 for the step after a refused command \\
    Max.\ output tokens per step & 8,192 \\
    Max.\ steps per run & 60 \\
    Verifier & CBMC 5.95.1, \texttt{--unwinding-assertions} \\
    Time limit per verifier call & 300\,s\\
    Repeated commands & refused after 5 repetitions \\
    \bottomrule
  \end{tabular}
\end{table}
The full configuration of our harness agent used in our experiments is shown in \Cref{tab:config}. The prompt and othe relevant artifacts are available on Zenodo \footnote{ https://doi.org/10.5281/zenodo.23378668}.

\subsection{Soundness Guard}

\begin{algorithm}[t]
\caption{\textsc{SoundnessGuard}}
\label{alg:harness-agent}
\begin{algorithmic}[1]
\Require Software system $S$, target function $f$,  proof harness $H_f$
\Ensure Proof harness $H_f$ for $f$

\State $D \gets \Call{Dependencies}{H_f}$
\ForAll{$d \in D$}
    \State $A_d \gets \Call{GetAbstraction}{H_f,d}$
    \State $\text{CEX}_d \gets \Call{Validate}{d,A_d}$
        \Comment{Check $d \Rightarrow A_d$}
    \While{$\text{CEX}_d \neq \emptyset$}
        \State $H_f \gets
            \Call{AgentRepair}{S,H_f,d,\text{CEX}_d}$
        \State $A_d \gets \Call{GetAbstraction}{H_f,d}$
        \State $\text{CEX}_d \gets \Call{Validate}{d,A_d}$
    \EndWhile
\EndFor
\State \Return $H_f$
\end{algorithmic}
\end{algorithm}



The soundness guard ensures that the agent does not introduce unsound restrictions when abstracting dependencies. For each dependency $d$ replaced by an abstraction $A_d$, we require that $A_d$ over-approximates the behavior of $d$ (written as $d \Rightarrow A_d$). This includes return values, but also the observable effect on the program state.

As shown in \Cref{alg:harness-agent}, the guard validates each proposed dependency abstraction against its original implementation. If validation reveals a behavior of $d$ excluded by $A_d$, the resulting counterexample $\text{CEX}_d$ is returned to the agent for repair. The revised abstraction is checked again before being accepted. Only successfully validated abstractions are incorporated into the harness.

In practice, validation is performed by a verification task: the agent generates a small program that calls the original~$d$ with arbitrary arguments and asserts~$A_d$ on its result, and runs the verifier on it. If the verifier finds arguments for which the assertion fails, $d$ can behave in a way that $A_d$ excludes and these arguments are returned to the agent as~$\mathit{CEX}_d$. If no such arguments are found, $A_d$ is accepted and recorded. The agent only uses the accepted abstractions in its harness.

Our validation approach relies on an initially sound over-approximation and preserves this property through checked refinements. Under these conditions, the soundness of the resulting harness does not depend on the correctness of the agent's reasoning.

\subsection{Precision Guard}

\begin{algorithm}[t]
\caption{\textsc{PrecisionGuard}}
\label{alg:proof-harness}
\begin{algorithmic}[1]
\Require Verifier $\mathcal{V}$, software system $S$, target function $f$, proof harness $H_f$  
\Ensure Verdict $v$, counterexample $\text{CEX}$ 


    \State $(\text{verdict}, \text{CEX}) \gets \mathcal{V}(H_f)$
    \If{$\text{verdict} = \textsc{Safe}$}
        \State \Return $(\textsc{Safe}, \emptyset)$
    \ElsIf{$\text{CEX} \neq \emptyset$}
        \If{$\Call{Feasible}{S,f,\text{CEX}}$}
            \State \Return $(\textsc{Unsafe}, CEX)$ \Comment{Feasible counterexample}
        \Else
            \State \Return $(\textsc{RefNeeded}, CEX)$
            \Comment{Refinement required}
        \EndIf
    \Else
        \State \Return $(\textsc{Unknown}, \emptyset)$ \Comment{Verifier returned unkown}
    \EndIf
\end{algorithmic}
\end{algorithm}




Soundness alone does not guarantee that a verifier can establish the desired property. An over-approximating harness may admit behaviors that cannot occur in the original system, potentially leading to spurious counterexamples. The precision guard therefore aims to refine the harness until it is sufficiently precise for the chosen verifier $\mathcal{V}$.

As shown in \Cref{alg:proof-harness}, the precision guard invokes $\mathcal{V}$ on the current harness $H_f$. If the verifier establishes the property, the process terminates successfully. Otherwise, a reported counterexample is checked by attempting to reproduce the violation through concrete execution of the target function. If the violation is reproduced, the counterexample is passed to the agent which reports it to the developer. Otherwise, the counterexample is returned to the harness agent to guide further refinement. Each proposed refinement is again subject to the soundness guard.

Importantly, this check does not establish whether the initial state of the counterexample is reachable in the original software system. We assume that developers encode the intended preconditions and correctness properties directly in the program. The harness agent therefore does not need to infer restrictions beyond those that real callers of the target visibly establish.

In our prototype, counterexamples are checked by replay. The tool extracts from the verifier's trace the arbitrary values used by the verifier. The agent then compiles the driver against the real program and runs it with these arbitrary values from the verifier. If the same assertion
fails, the violation is confirmed and reported. If the program runs without failure, the counterexample is spurious i.e. it relied on behaviour that only the over-approximated abstraction allows.

Conversely, when the verifier reports that no violation exists, the tool checks that at least one execution of the harness reaches the end of the created driver. Otherwise, the assumptions exclude all executions and the result would be vacuous, such results are rejected.

The process terminates when the verifier establishes the property, a counterexample is reproduced, or verification remains inconclusive. Thus, the harness need not represent the execution environment exactly; it only needs to be sufficiently precise for the verifier to establish a conclusive result.

\section{Example of Abstraction Refinements}\label{sec:refinement}

\begin{figure}[t]
\begin{minipage}[t]{.35\textwidth}
\begin{lstlisting}[style=figcode]
// REAL FUNCTION
short toAdd(short val) {
  #if VAL_POS
    if (val < 0) {
        val = -val;
    }
  #endif
  if (val < -RANGE) {
        return -RANGE;
  } else if (val > RANGE) {
        return RANGE;
  }
  return val;
}
\end{lstlisting}
\end{minipage}
\hfill
\begin{minipage}[t]{.55\textwidth}
\begin{lstlisting}[style=figcode]
// 1st abstraction
short toAdd(short val) {
    return nondet_short();
}
// 2nd abstraction 
short toAdd(short val) {
    short r = nondet_short();
    assume(-RANGE <= r && r <= RANGE);
    return r;
}
// 3rd abstraction
short toAdd(short val) {
    short r = nondet_short();
    assume(r >= -RANGE && r <= RANGE && 
          (r >= 0 $||$ r == -RANGE));
    return r;
}
// 4th abstraction 
short toAdd(short val) {
    short r = nondet_short();
    assume((r >= 0 && r <= RANGE)  
        $||$ (val == SHRT_MIN 
              && r == -RANGE));
    return r;
}
\end{lstlisting}
\end{minipage}
\caption{Function \texttt{toAdd}  and its abstractions in different refinements }
\label{fig:mock}
\end{figure} 

In \Cref{fig:mock}, we give the abstractions of the dependencies of target function $f=\texttt{add}$ of \Cref{fig:example} while constructing $H_f$. Function \texttt{add} calls one function \texttt{toAdd} given in \Cref{fig:mock} (left). Function \texttt{toAdd} is supposed to bring the input value \texttt{val} into the range \texttt{[-RANGE,+RANGE]} plus make it positive when the build flag \texttt{VAL\_POS} is set to 1. It, however, does not correctly achieve this because of an overflow when \texttt{val} equals \texttt{SHRT\_MIN}. As a consequence, the assertion in target function \texttt{add} of \Cref{fig:example} (left) is violated. 

For constructing the harness $H_f$ for a makefile with \texttt{VAL\_POS=1} and \texttt{RANGE=10}, the harness agent starts with an overapproximation of \texttt{toAdd} just specifying \texttt{toAdd} to return a value of type \texttt{short} (see \cref{fig:mock} (right), 1st abstraction). Step \textsc{Validate} in \Cref{alg:harness-agent} sees this to be a valid overapproximation and does not enter its refinement loop. When invoking the verifier $\mathcal{V}$ on the current $H_f$ in the precision guard (\Cref{alg:proof-harness}), it however finds a counterexample (invocation with \texttt{val=0}, return value -67)  leading to a violated assertion. This counterexample is, however, not feasible (in an execution of \texttt{add}  with this input value using the real function \texttt{toAdd}, the assertion in the target function is not violated). The agent thus refines the abstraction of  \texttt{toAdd} (see  \cref{fig:mock} (right), 2nd abstraction). This abstraction again passes the soundness guard, but not the precision guard which produces another spurious counterexample (\texttt{val=0}, return -4).  A further refinement produces the 3rd abstraction, again being sound, but not passing the precision guard (counterexample \texttt{val=0}, return -10). The fourth abstraction now passes the precision guard with a feasible counterexample (\texttt{val=SHRT\_MIN}, return -10) and this completes the construction of harness $H_f$. The agent thus reports \textsc{Unsafe}   and also gives harness and feasible counterexample to a developer for further inspection. 

Note that in our example here, the soundness guard never needs to refine the abstraction; all abstractions are valid overapproximations of \texttt{toAdd}. In our initial experiments, we have however seen that some LLMs do provide wrong abstractions, making the usage of a soundness guard mandatory for sound harnesses.

\section{Minimal Verifier-Specific Harness Adaption}\label{app:harness}

Our agent supports multiple harness formats per verification task. Importantly, the agent does not produce verifier-specific harnesses, but provides us with sufficient context to construct them deterministically (e.g. target file, additional source files, dependency abstractions, and build flags). Therefore, the same harness can be reused with multiple verifiers by using minimal deterministic adaptions. For CBMC, used inside the agent's loop, the harness additionally records every value of every symbolic input, so that counterexamples can be replayed. For CPAchecker and Ultimate Automizer, the tool generates an SV-COMP-style harness. It uses the standard \texttt{\_\_VERIFIER\_*} functions and a failing assertion calls \texttt{reach\_error}. The program is preprocessed with GCC into a single file. For KLEE, the same harness is mapped to KLEE's functions and preprocessed with clang. All tasks therefore share the same driver, target code and dependency abstractions. They only differ in how they express symbolic inputs, assumptions and failures.

\section{Prompt}\label{sec:prompt}

The prompt in Listing~\ref{lst:prompt} summarises the harness format and the guidance the agent receives to generate a harness. The agent's overall behaviour, including the task description and the rules it must follow, is set by a system prompt. The full prompt is part of the artifact.

\lstset{basicstyle=\ttfamily\fontsize{5.5}{6.2}\selectfont, breaklines=true,
        columns=fullflexible, keepspaces=true, frame=single, numbers=none,
        xleftmargin=0pt, aboveskip=2pt, belowskip=2pt}

\begin{lstlisting}[caption={Prompt given to the agent to generate the harness (excerpt).},
                   label={lst:prompt}]
Write two files in <project>/verification/<target>/:

spec.json   target and its source file; extra source files to include with their
            real code; headers; build flags; how to link the real program for
            replay; loop bound; mocks. [...]
driver.c    a small program that calls the target the way a real caller would.

Driver
- Inputs that real callers may vary: unknown values, __VERIFIER_nondet_<type>().
  A fixed input checks only that one input and proves nothing.
- Guarantees that real callers provide: __VERIFIER_assume(<condition>).
- Keep inputs small, so that every loop fits within the bound.
- Derive the calling context from existing callers and tests.
- Do not add assertions; only the project's own assertions are checked.
[...]

Dependencies
- Functions without code in the harness return ANY value. If such a function
  matters, include its source file or mock it.
- Mocks are allowed only for functions defined outside the target's file, and only
  for number-valued parameters and return values. A mock returns an unknown value,
  optionally restricted by a rule: a C condition over the return value r and the
  parameter names.

Refinement (given through the hints after each tool result)
- Spurious counterexample: the hint shows the arguments a mock received, the value
  it returned, and the value the real function returns. Propose a rule that
  excludes the mock's value.
- Before use, every rule must pass `check` against the real function.
  Refuted: the hint shows an input where the real function breaks the rule;
  propose a weaker rule. Not refuted: add the rule to the spec and run again.
[...]

Finish with a verdict reported by a tool.
\end{lstlisting}

\end{document}